\documentclass[prd,floatfix,notitlepage,noshowpacs,noshowkeys,nofootinbib,superscriptaddress]{revtex4-1}
\usepackage{natbib}
\usepackage{times}
\usepackage{amssymb,amsbsy,amsmath,amsfonts}
\usepackage{graphicx}
\usepackage{epstopdf}
\usepackage{float}
\usepackage{color}
\usepackage{morefloats}
\usepackage{rotating}
\usepackage{srcltx}
\usepackage{slashed}
\usepackage{subfigure}
\usepackage{multirow}
\usepackage{verbatim}
\usepackage{hyperref}
\usepackage{tabularx}

\graphicspath{{./}{./img/}{./fig/}{./image/}{./figure/}{./picture/}}

\begin{document}

\title{Antineutrino induced $\mathbf{\Lambda(1405)}$ production off the proton}
\author{Xiu-Lei Ren}
\email[E-mail: ]{xiulei.ren@buaa.edu.cn}
\affiliation{School of Physics and
Nuclear Energy Engineering \& International Research Center for Nuclei and Particles in the Cosmos, Beihang University, Beijing 100191, China}
\affiliation{Departamento de F\'{\i}sica Te\'{o}rica and IFIC,
Centro Mixto Universidad de Valencia-CSIC, Institutos de Investigaci\'{o}n de Paterna,
Apartado 22085, 46071 Valencia, Spain}
\affiliation{Institut de Physique Nucl\'{e}aire, IN2P3-CNRS and Universit\'{e} Paris-Sud,
F-91406 Orsay Cedex, France}

\author{E. Oset}
\email[E-mail: ]{oset@ific.uv.es}
\affiliation{Departamento de F\'{\i}sica Te\'{o}rica and IFIC,
Centro Mixto Universidad de Valencia-CSIC, Institutos de Investigaci\'{o}n de Paterna,
Apartado 22085, 46071 Valencia, Spain}

\author{L. Alvarez-Ruso}
\email[E-mail: ]{alvarez@ific.uv.es}
\affiliation{Instituto de F\'isica Corpuscular (IFIC),
Centro Mixto Universidad de Valencia-CSIC, Institutos de Investigaci\'{o}n de Paterna,
Apartado 22085, 46071 Valencia, Spain}

\author{M. J. Vicente Vacas}
\email[E-mail: ]{vicente@ific.uv.es}
\affiliation{Departamento de F\'{\i}sica Te\'{o}rica and IFIC,
Centro Mixto Universidad de Valencia-CSIC, Institutos de Investigaci\'{o}n de Paterna,
Apartado 22085, 46071 Valencia, Spain}

\begin{abstract}
 We have studied the strangeness changing antineutrino induced reactions $\bar{\nu}_{l} p \rightarrow l^+ \phi B $, with $\phi B = K^-p$, $\bar{K}^0n$, $\pi^0\Lambda$, $\pi^0\Sigma^0$, $\eta\Lambda$, $\eta\Sigma^0$, $\pi^+\Sigma^-$, $\pi^-\Sigma^+$, $K^+\Xi^-$ and $K^0\Xi^0$, using a chiral unitary approach. These ten coupled channels are allowed to interact strongly, using a kernel derived from the chiral Lagrangians.  This interaction generates two $\Lambda(1405)$ poles, leading to a clear single peak in the $\pi \Sigma$ invariant mass distributions. At backward scattering angles in the center of mass frame, $\bar{\nu}_{\mu} p \rightarrow \mu^+ \pi^0 \Sigma^0$  is dominated by the $\Lambda(1405)$ state at around 1420~MeV while the lighter state becomes relevant as the angle decreases, leading to an asymmetric line shape. In addition, there are substantial differences in the shape of $\pi \Sigma$ invariant mass distributions for the three charge channels. If observed, these differences would provide valuable information on a claimed isospin $I=1$, strangeness $S=-1$ baryonic state around $1400$~MeV. Integrated cross sections have been obtained for the $\pi \Sigma$ and $\bar K N$ channels, investigating the impact of unitarization in the results. The number of events with $\Lambda(1405)$ excitation in $\bar\nu_\mu p$ collisions in the recent antineutrino run at the MINER$\nu$A experiment has also been obtained. We find that this reaction channel is relevant enough to be investigated experimentally and to be taken into account in the simulation models of future experiments with antineutrino beams.
\end{abstract}



\maketitle
\section{Introduction}

The $\Lambda(1405)$ resonance is a cornerstone in hadron physics, 
challenging the standard view of baryons made of three quarks.
Long ago it was already suggested that the $\Lambda(1405)$ could be a kind
of molecular state arising from the interaction of the $\pi\Sigma$ and $\bar{K}N$
channels~\cite{dalitz,dalitzdos}. This view has been recurrent~\cite{tonythomas}, but
only after the advent of unitary chiral perturbation theory (UChPT) has it taken 
a more assertive tone~\cite{weise,angels,cola,ollerulf,hyodo,carmen}. In 
this framework, a kernel (potential) derived from the chiral Lagrangians 
is the input into the Bethe-Salpeter equation in coupled channels. Sometimes the
interaction is strong enough to generate poles, denominated as dynamically 
generated states, which can be interpreted as hadronic molecules 
with components on the different channels (see Ref.~\cite{review} for a review).

It came as a surprise that UChPT predicts two $\Lambda(1405)$ states~\cite{ollerulf}, 
studied in detail in Ref.~\cite{cola}. Two poles appear, one around 1420~MeV with a 
width of about 40~MeV and another one around 1385~MeV with a larger width of about 150~MeV. 
These findings have been reconfirmed in more recent studies with potentials that include higher order
terms of the chiral Lagrangians~\cite{borasoyweise, borasoymeissner,hyodoweise,Feijoo:2015yja,kanchan,hyodorev,ollerguo}. From the experimental perspective, the old experiments~\cite{thomas, hemingway} produced $\pi\Sigma$ invariant mass distributions where a single $\Lambda(1405)$ peak is seen around $1405$~MeV. According to Ref.~\cite{osethyodo}, this single peak results from the overlap 
of the two pole contributions. It has also been suggested that reactions induced by $K^- p$ pairs show a peak around 1420~MeV because the pole at 1420~MeV couples mostly to $\bar{K}N$, while the one at $1385$ MeV does it more strongly to $\pi\Sigma$. This would be the case of $K^-p\rightarrow \gamma \pi \Sigma$~\cite{osetramosjc} and $K^-p\rightarrow \pi^0\pi^0\Sigma^0$. The latter one, measured at Crystal Ball~\cite{nefkens} and analyzed in Ref.~\cite{angelsmagas}, confirmed the existence of the state at $1420$~MeV. Another reaction that has proved its existence is $K^- d \rightarrow n \pi \Sigma$~\cite{braun}, which was studied in Ref.~\cite{sekihara}. The issues raised in Ref.~\cite{miyagawa} were addressed in detail in Ref.~\cite{sekidos} reconfirming the findings of Ref.~\cite{sekihara}.

It is somewhat surprising that the two poles emerge in the theory even when only data on $K^-p$ scattering and $K^-p$ atoms~\cite{sidharta}, which are above the $\Lambda(1405)$ pole masses, are fitted.  Nevertheless, it is clear that the best information on the $\Lambda(1405)$ properties should come from processes where the $\Lambda(1405)$ is produced close to its pole masses. In this sense, the abundant $\Lambda(1405)$ photoproduction data obtained by CLAS with the $\gamma p\rightarrow K^+\pi^+\Sigma^-$, $K^+\pi^0\Sigma^0$, $K^+\pi^-\Sigma^+$ reactions~\cite{moriya} add much information to the earlier data of Ref.~\cite{niiyama}, bringing  new light into the subject. A fit to these data imposing unitarity in the $\pi\Sigma, \bar{K}N$ channels and allowing only small variations in the kernel of the chiral Lagrangians~\cite{roca,rocados} has reconfirmed the existence of the two poles, in agreement with the UChPT predictions. The wide range of energies investigated and the simultaneous measurement of the three $\pi\Sigma$ charged channels were the key to the solutions found in Refs.~\cite{roca,rocados} and, more recently, in Ref.~\cite{mai}.

Studies of $p \, p  \rightarrow p \, K^+ \, \Lambda(1405)$ performed at ANKE show again a superposition of the contributions from the two poles~\cite{anke}, and can be explained with the theoretical framework of UChPT~\cite{genganke}. More recent measurements~\cite{fabbietti,fabbidos} show the $\Lambda(1405)$ peak at a lower energy than in the ANKE experiment~\cite{anke}. Some reasons for this behavior have been suggested in Ref.~\cite{fabbidos}. If more data for this reaction on different conditions became available, a global analysis like the one of Ref.~\cite{roca,rocados} for photoproduction would be advisable. In between, $\Lambda(1405)$  electroproduction~\cite{electro} data [$e \, p \rightarrow e' K^+ \, \Lambda(1405)$] have unexpectedly revealed a two-peak structure, albeit with large uncertainties. Previous measurements with different reactions have only observed a single peak coming from the superposition of the two poles, with different shapes depending on the weight of either pole, as determined by the dynamics of each process. 

Lattice QCD simulations have also brought new light into the $\Lambda(1405)$ properties. Using three-quark interpolators, a state associated with the $\Lambda(1405)$  is produced~\cite{leinweber,lang}. 
The vanishing strange quark contribution to the $\Lambda(1405)$ magnetic moment for light quark masses close to the physical ones has been interpreted~\cite{Hall:2014uca,ross} as an evidence of a large $\bar K N$ component in the wave function of the $\Lambda(1405)$.  Further work along these lines was reported in Ref.~\cite{albermela} using {\it synthetic} lattice results from $\bar K N$ and $\pi \Sigma$ interpolators. These lead to the right description of the meson-baryon amplitudes in the continuum and contain the two poles in the complex plane.

Until now, the weak excitation of $\Lambda(1405)$ has never been investigated. It is remarkable that while its production in strong and electromagnetic processes has to involve an extra strange particle (usually a $K^-$ in the initial state or a $K^+$ in the final one), the direct excitation of $\Lambda(1405)$ induced by antineutrinos $\bar\nu_l p \rightarrow l^+ \Lambda(1405)$ is allowed although Cabibbo suppressed. Notice that in $\Lambda(1405)$ photo and electroproduction there are line shape distortions due to final state interactions between the $K^+$ and the $\Lambda(1405)$ decay products, which are absent in the weak reaction.

Stimulated by the precision needs of neutrino oscillation experiments, there is a significant ongoing effort aimed at a better understanding of neutrino cross sections with nucleons and nuclei. The goal is to develop better interaction models to reduce systematic errors in the detection process, constrain irreducible backgrounds and achieve a better neutrino energy determination.\footnote{Neutrino beams are not monochromatic so that the incident energy is not known for single events. However, oscillation probabilities are functions of this a priori unknown quantity.} In the recent past, several experiments have produced valuable cross section measurements (see Ref.~\cite{Formaggio:2013kya} for a comprehensive review of the available data). The MINER$\nu$A experiment~\cite{Minerva,Drakoulakos:2004gn} at FNAL, fully dedicated to the study of neutrino interactions with different target materials has recently completed data taking and started to produce interesting results~\cite{Fields:2013zhk,Fiorentini:2013ezn,Tice:2014pgu,Higuera:2014azj}.

In the few-GeV energy region, where several of the current and future experiments operate, quasielastic scattering and single pion production have the largest cross sections but strange particle production is also relevant. The charged-current $\Delta S =-1$ quasielastic hyperon ($Y =\Lambda,\Sigma$) production by antineutrinos has been investigated~\cite{Singh:2006xp,Mintz:2007zz,Kuzmin:2008zz} and found to be a non-negligible source of pions through the $Y \rightarrow N \, \pi$ decay~\cite{Singh:2006xp,Alam:2013cra}. 
Among the inelastic processes, associated ($\Delta S =0$) production of $\bar K$ and $\Sigma$ or $\Lambda$ baryons is the dominant one but has a high threshold. Below it, single $K$ ($\Delta S =1$) and single $\bar K$ ($\Delta S =-1$) can be produced in charged current interactions induced by $\nu$ and $\bar \nu$ respectively. These processes have been recently studied using SU(3) chiral Lagrangians at leading order~\cite{RafiAlam:2010kf,manolok}. The weak hadronic currents and the corresponding cross sections at threshold are constrained by chiral symmetry with couplings extracted from pion and hyperon semileptonic decays. As stressed in Ref.~\cite{Alvarez-Ruso:2014bla}, while the derived $K$ production cross section is a robust prediction at threshold, the situation could be different for $\bar K$ production due to the presence of the  $\Lambda(1405)$ resonance just below the $\bar K N$ threshold. Another, so far unexplored, $\Delta S = -1$ reaction that can occur below the associated production threshold, $\bar\nu_l \, p \rightarrow l^+ \, \Sigma \, \pi$, is bound to get an important contribution from $\Lambda(1405)$ excitation.    

Here we report the first study of the antineutrino induced reactions $\bar{\nu}_l p \rightarrow l^+ \phi B $ with $\phi B =$ $K^-p$, $\bar{K}^0n$, $\pi^0\Lambda$, $\pi^0\Sigma^0$, $\eta\Lambda$, $\eta\Sigma^0$, $\pi^+\Sigma^-$, $\pi^-\Sigma^+$, $K^+\Xi^-$, $K^0\Xi^0$ in coupled channels, paying special attention to the role of the $\Lambda(1405)$. In Sect.~\ref{SecII} we describe the theoretical framework. The results are presented in Sect.~\ref{SecIII} followed by our conclusions.

\section{Theoretical Framework}
\label{SecII}

\subsection{Effective Lagrangians}

\begin{figure}[h!]
  \centering
  \includegraphics[width=\textwidth]{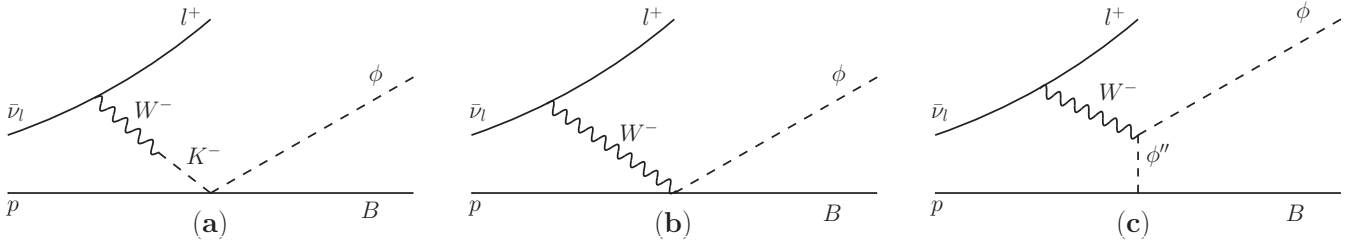}
  \caption{Feynman diagrams for the process $\bar{\nu}_l p \rightarrow l^+ \phi B$. (a) denotes the kaon pole term (KP), (b) represents the contact term (CT), and (c) stands for the meson ($\phi''$) in-flight term (MF).}
  \label{Fig:tree}
\end{figure}

At tree level, the process $\bar{\nu}_l p \rightarrow l^+ \phi B$, with $\phi$ and $B$ being the meson and baryon in the final state, proceeds as depicted in the diagrams of Fig.~\ref{Fig:tree}. There are also baryon-pole terms (see Fig.~1 of Ref.~\cite{manolok}) which contribute predominantly to the $p$-wave state of the $\phi B$ system. Since our aim is to generate the $\Lambda(1405)$, which appears in $\phi B$ s-wave, we neglect these terms. 

All mechanisms in Fig.~\ref{Fig:tree} consist of a leptonic and a hadronic currents that interact via the exchange of a $W$ boson. The leptonic part is provided by the Standard Model Lagrangian
\begin{equation}\label{Eq:LL}
   \mathcal{L}=-\frac{g}{2\sqrt{2}} \left[ \bar{\psi}_\nu \gamma_\mu (1-\gamma_5) \psi_l W^\mu + \bar{\psi}_l \gamma_\mu (1-\gamma_5) \psi_\nu {W^{\dag}}^\mu\right] \,,
\end{equation}
where $\psi_\nu$, $\psi_l$ and $W$ denote the neutrino, charged lepton and gauge boson $W$ fields, respectively;  $g$ is the gauge coupling, related to the Fermi constant by $G_F=\sqrt{2} g^2/(8M_W^2) = 1.16639(1)\times 10^{-5}$ GeV$^{-2}$.

The hadronic current is derived from chiral Lagrangians~\cite{gasser,pich,Scherer:2002tk} at leading order. As mentioned above, in this work  we are only concerned about the $s$-wave contribution. In the meson sector, required for CT and MF diagrams, the lowest order SU(3) Lagrangian is given by
\begin{equation}
  \mathcal{L}_\phi^{(2)} = \frac{F_0^2}{4}\langle D_\mu U (D^\mu U)^\dag\rangle + \frac{F_0^2}{4}\langle \chi U^\dag + U\chi^\dag\rangle,
\end{equation}
where $\langle \ldots \rangle$ stands for the trace in flavor space; $F_0$ is the pseudoscalar meson decay constant in the chiral limit. The quantity $\chi=2B_0\mathcal{M}$, with the quark-mass matrix $\mathcal{M} = \mathrm{diag}(m_u,m_d,m_s)$, represents the explicit breaking of chiral symmetry. The function $U = \mathrm{exp}\left(i \phi/F_0\right)$ is the SU(3) representation of the meson fields 
\begin{equation}
\phi = \left(\begin{array}{ccc}
\pi^0 + \frac{1}{\sqrt{3}} \eta & \sqrt{2} \pi^+ & \sqrt{2} K^+ \\
\sqrt{2} \pi^- &  -\pi^0 + \frac{1}{\sqrt{3}} \eta & \sqrt{2} K^0 \\
\sqrt{2} K^- & \sqrt{2} \bar{K}^0 & -\frac{2}{\sqrt{3}} \eta 
\end{array}\right) \,,
\end{equation}  
and its covariant derivative $D_\mu U$ can be written as
\begin{equation}
  D_\mu U = \partial_\mu U - i r_\mu U + i U l_\mu,
\end{equation}
where $l_\mu$ and $r_\mu$ correspond to left- and right-handed currents. For the charged current weak interaction 
\begin{equation}
  r_\mu = 0,\quad l_\mu = \frac{g}{\sqrt{2}} ( W_\mu^\dagger T_+ + W_{\mu} T_-),
\end{equation}
with
\begin{equation}
  T_+ = \left(\begin{array}{ccc}
          0 & V_{ud} & V_{us} \\
          0 & 0 & 0 \\
          0 & 0 & 0
        \end{array}\right), \quad
  T_- = \left(\begin{array}{ccc}
          0 & 0 & 0 \\
          V_{ud} & 0 & 0 \\
          V_{us} & 0 & 0
        \end{array}\right).
\end{equation}
Here, $V_{ij}$ are the relevant elements of the Cabibbo-Kobayashi-Maskawa matrix. Their magnitudes are $|V_{ud}|=\cos \theta_c =0.97425\pm0.00022$ and $|V_{us}|= \sin \theta_c =0.2252\pm0.0009$~\cite{pdg}, with $\theta_c$ the Cabibbo angle.

The lowest order chiral effective Lagrangian describing the interaction between the octet of pseudoscalar mesons and the octet of baryons can be written as
\begin{equation}~\label{Eq:MBL}
  \mathcal{L}_{\phi B}^{(1)} = \langle\bar{B}(i\slashed D-M_B)B\rangle+\frac{D}{2}
    \langle\bar{B}\gamma^{\mu}\gamma_5\{u_{\mu},B\}\rangle +\frac{F}{2}
    \langle\bar{B}\gamma^{\mu}\gamma_5[u_{\mu},B]\rangle \,,
\end{equation}
with the baryon fields arranged in the matrix
\begin{equation}
B = \left(\begin{array}{ccc}
\frac{1}{\sqrt{2}} \Sigma^0 + \frac{1}{\sqrt{6}} \Lambda &  \Sigma^+ & p \\
\Sigma^- & -\frac{1}{\sqrt{2}} \Sigma^0 + \frac{1}{\sqrt{6}} \Lambda  & n \\
\Xi^- & \Xi^0 &  -\frac{2}{\sqrt{6}} \Lambda
\end{array}\right) \,;
\end{equation}
$M_B$ denotes the baryon octet mass in the chiral limit; $D=0.804$ and $F=0.463$ are the axial-vector coupling constants, which are determined from the baryon semi-leptonic decays~\cite{Cabibbo:2003cu}. The covariant derivative of the baryon field is defined as
\begin{equation}
  D_{\mu}B=\partial_{\mu}B+[\Gamma_{\mu} , B],
\end{equation}
\begin{equation}
  \Gamma_{\mu}=\frac{1}{2}\left\{u^{\dag}(\partial_{\mu}-i r_{\mu})u + u(\partial_{\mu}-i l_{\mu})u^{\dag}\right\},
\end{equation}
and $u_{\mu}$ is given by
\begin{equation}
  u_{\mu}=i\left\{u^{\dag}(\partial_{\mu}-i r_{\mu})u-u(\partial_{\mu}-i l_{\mu})u^{\dag}\right\} ,
\end{equation}
where $u=\sqrt{U}$.

\subsection{Chiral Unitary Theory}
\label{subsec:UChPT}

\begin{figure}[h!]
  \centering
  \includegraphics[width=\textwidth]{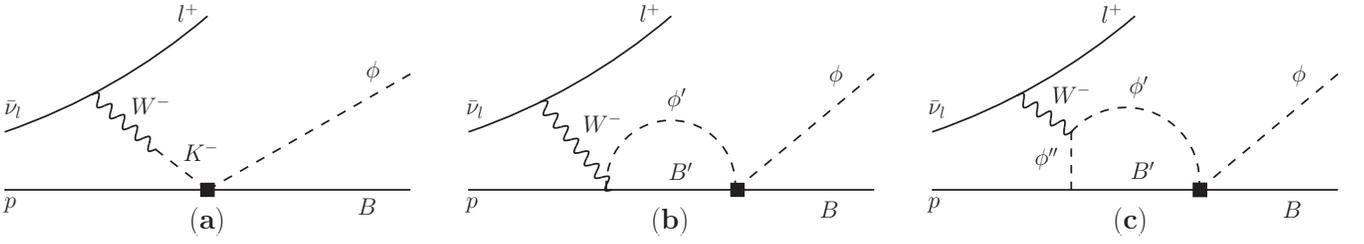}\\
  \caption{Iterated loop diagrams for $\bar{\nu}_\mu p \rightarrow \mu^+ \phi B$. The solid boxes represent the $T$ matrix of the ten coupled channels.}
  \label{Fig:loop}
\end{figure}

As discussed in the introduction, the $\Lambda(1405)$ is dynamically generated by the interaction of $S=-1$ $s$-wave meson-baryon pairs in coupled channels. This can be achieved by solving the Bethe-Salpeter equation with the interaction potential  provided by the chiral Lagrangian of Eq.~(\ref{Eq:MBL}). In the diagrams of Fig.~\ref{Fig:tree}, the outgoing meson and baryon can interact producing the resonance. Therefore, one must consider the diagrams depicted in Fig.~\ref{Fig:loop}. The solid square in the figures represents the different $T_{ij \rightarrow \phi B}$ amplitudes, where the pair of indices $i j = K^-p, \, \bar{K}^0n, \, \pi^0\Lambda,\, \pi^0\Sigma^0,\, \eta\Lambda,\, \eta\Sigma^0,\, \pi^+\Sigma^-,\, \pi^-\Sigma^+,\, K^+\Xi^-,\, K^0\Xi^0$ denote any of the ten allowed channels.

Following the approach of Ref.~\cite{angels} for the strong interaction in the $S=-1$ sector,
\begin{equation}
  T = V + VG T  = [1-VG]^{-1}V \,,
\label{Eq:BS}
\end{equation}
where the lowest-order interaction amplitude $V$, extracted from the lowest order chiral Lagrangian $\mathcal{L}^{(1)}_{\phi B}$, is given by
\begin{equation}
  V_{ij} = -C_{ij}\frac{1}{4F_{\phi}^2}(k^0 + {k'}^0) 
\end{equation}
after a nonrelativistic reduction. Here, $k^0$ and ${k'}^{0}$ are the energies of the incoming and outgoing mesons in the $\phi B$ center of mass (CM) frame; $F_0$ has been replaced by the average value of the physical decay constants $F_{\phi}=1.15 f_{\pi}$ with $f_\pi=93$ MeV as in Ref.~\cite{angels}. The $10 \times 10$ matrix of coefficients $C_{ij}$ can be found in Table~1 of Ref.~\cite{angels}.

The meson-baryon loop function $G_{ij}$ is given by
\begin{eqnarray}
  G_{ij} &=& i\int \frac{d^4 q}{(2\pi)^4} \frac{M_j}{E_j(\vec{q\ })} \frac{1}{k^0+p^0-q^0-E_j(\vec{q\ }) + i\epsilon} \frac{1}{q^2 - m_i^2 + i\epsilon}\,,\nonumber\\
  &=& \int \frac{d^3 q}{(2\pi)^3} \frac{1}{2\omega_{j}(\vec{q\ })}\frac{M_j}{E_j(\vec{q\ })}\frac{1}{p^0+k^0 -\omega_i(\vec{q\ })-E_j(\vec{q\ })+i\epsilon}\,,
\end{eqnarray}
where $m_i$, $M_j$ are the physical meson and baryon masses of the $ij$ state while $\omega_i = (m_i^2+\vec{q}^{\,2})^{1/2}$, $E_j=(M_j^2+\vec{q}^{\,2})^{1/2}$ are the corresponding energies. It is a function of the CM energy $M_\mathrm{inv} = p^0+k^0$. In Ref.~\cite{angels},  the loop function is regularized with a cutoff $q_\mathrm{max}=630$~MeV.

\subsection{Cross section}

The reaction under consideration is 
\begin{equation}\label{Eq:reaction}
  \bar{\nu}_l(k_{\bar{\nu}}) + p(p) \rightarrow l^{+}(k_l) + \phi(k^{\prime}) + B(p^{\prime}),
\end{equation}
where $k_{\bar{\nu}}=(k^0_{\bar\nu},\vec{k}_{\bar\nu})$ [$k_l=(k^0_l,\vec{k}_l)$] is the 4-momentum of the incoming neutrino [outgoing charged lepton] while $p=(E_p,\vec{p})$, $p'=(E_B,\vec{p}\,')$ and $k'=(\omega_\phi,\vec{k}')$ denote the momenta of the initial proton, final baryon and final meson, in this order. Its cross section is given by
\begin{equation}
\sigma = \frac{2 M_p m_{\bar\nu}}{\lambda^{1/2}(s,m^2_{\bar\nu},M_p^2)} \int\frac{d^3k_l}{(2\pi)^3} \frac{m_l}{k^0_l}\int\frac{d^3k'}{(2\pi)^3}\frac{1}{2\omega_\phi} \int\frac{d^3p'}{(2\pi)^3}\frac{M_B}{E_B}
(2\pi)^4\delta^4(p+k_{\bar{\nu}}-k_l-k'-p')\overline{\sum} |t|^2 \,,
\end{equation}
where $\lambda(x,y,z) = x^2 + y^2 + z^2 - 2 x y - 2 x z - 2 y z$ and $s=(p+k_{\bar\nu})^2$; $\overline{\sum}$ denotes the sum over final state polarizations and average over the initial ones. It is convenient to perform the integrals over $\vec{p}\,'$ and $\vec{k}'$ in the $\phi B$ CM frame, taking advantage of the fact that the amplitude is projected onto the $s$-wave state of the $\phi B$ pair. The last integration over $\vec{k}_l$ is carried out in the global ($\bar\nu p$)  CM frame. We obtain
\begin{equation}\label{Eq:sigma}
  \sigma = \frac{2}{(2\pi)^3}\frac{m_{\bar{\nu}} m_l M_p M_B}{\sqrt{s}(s-M_p^2)} \int_{m_\phi + M_B}^{\sqrt{s}-m_l} dM_\mathrm{inv}\int_{-1}^{+1} d\cos\theta |\vec{k}_l|_{\bar{\nu} p} |\vec{k}'|_{\phi B}\overline{\sum} |t|^2 \,,
\end{equation}
where $\theta$ is the angle between $\vec{k}_l$ and $\vec{k}_{\bar\nu}$ in the $\bar\nu p$ CM frame. In Eq.~(\ref{Eq:sigma})
\begin{equation}
  |\vec{k}_l|_{\bar{\nu}_l p} = \frac{\lambda^{1/2}(s, m_l^2, M_\mathrm{inv}^2)}{2\sqrt{s}}, \quad
  |\vec{k}'|_{\phi B} = \frac{\lambda^{1/2}(M_\mathrm{inv}^2, m_\phi^2, M_B^2)}{2M_\mathrm{inv}} 
\end{equation}
are the charged-lepton momentum in the $\bar\nu p$ CM frame and the meson momentum in the $\phi B$ CM frame, respectively.

\subsection{Invariant amplitude}

In the $(k_l-k_{\bar\nu})^2 \equiv q^2 \ll M_W^2$ limit, the amplitude can be cast as 
\begin{equation}
  -it = 2G_F V_{us} L^{\mu} H_{\mu} \,,
\end{equation}
where the leptonic current is 
\begin{equation}
L^{\mu}=\bar{v}(k_{\bar\nu})\gamma^{\mu}(1-\gamma_5)v(k_l) \,,
\end{equation}
while the hadronic current 
\begin{equation}
H_{\mu} = \bar{u}(p')\Gamma_\mu u(p)
\end{equation}
is determined by the sum of the following contributions
\begin{itemize}
\item KP (vector)
\begin{equation}
  \Gamma_{\mu}^\mathrm{KP} = -\frac{1}{2}F_{\phi}\frac{q_{\mu}}{q^2-m_{K^-}^2+i\epsilon} T_{K^-p\rightarrow \phi B}\,.
\end{equation}
Note that in Fig.~\ref{Fig:loop}~(a), the sum over the intermediate states $\phi' B'$ produces the $K^- p \rightarrow \phi B$ $t$-matrix element by virtue of Eq.~\ref{Eq:BS}.

\item CT (vector plus axial)
\begin{eqnarray}
  \Gamma_{\mu}^{\mathrm{CT}(V)} &=& -\frac{1}{4F_{\phi}}\left[C_{\phi B}^{(V)} \gamma_\mu + \sum\limits_{\phi' B'} C_{\phi' B'}^{(V)} \gamma_{\mu} G_{\phi'B'} T_{\phi'B'\rightarrow \phi B}\right],\label{Eq:CT1}\\
  \Gamma_{\mu}^{\mathrm{CT}(A)} &=& -\frac{1}{4F_{\phi}}\left[C_{\phi B}^{(A)} \gamma_\mu\gamma^5 + \sum\limits_{\phi' B'} C_{\phi' B'}^{(A)} \gamma_{\mu} \gamma^5 G_{\phi'B'} T_{\phi'B'\rightarrow \phi B}\right].\label{Eq:CT2}
\end{eqnarray}
The coefficients  $C_{\phi B}^{(V)}$ and $C_{\phi B}^{(A)}$ are tabulated in Table~\ref{Tab:PB1} and Table~\ref{Tab:PB2}, respectively. The loop function is given by
\begin{equation}
  G_{\phi'B'} = i\int\frac{d^4l}{(2\pi)^4}\frac{1}{l^2-m_{\phi'}^2+ i\epsilon} \frac{1}{\slashed p + \slashed q-\slashed l - M_{B'}+i\epsilon}.
\end{equation}

\item MF (axial)
\begin{equation}\label{Eq:MF}
  \Gamma_{\mu}^\mathrm{MF} = \frac{1}{4\sqrt{2} F_\phi}\left[\sum\limits_{\phi''} C_{\phi''\phi}C_{\phi''B}\frac{(2k'-q)_{\mu}(k'-q)_{\nu}\gamma^{\nu}\gamma^5} {(k'-q)^2-m_{\phi''}^2+i\epsilon} + \sum\limits_{\phi'\phi''B'} C_{\phi''\phi'}C_{\phi''B'} G_{\phi'\phi''B'}^\mu T_{\phi'B'\rightarrow \phi B}\right],
\end{equation}
where $\phi''$ denotes the internal meson in the tree level diagram (c) of Fig.~\ref{Fig:tree}. In most cases, only one type of meson can be exchanged but it happens that both $\pi^0$ and $\eta$ are allowed intermediate states. The $G^{\mu}_{\phi'\phi''B'}$ function is given by
\begin{equation}
  G_{\phi'\phi''B'}^\mu = i\int\frac{d^4 l}{(2\pi)^4}~ (2l-q)^\mu ~(l-q)^\nu ~\gamma_{\nu}\gamma^5  
\frac{1}{l^2-m_{\phi'}^2+i\epsilon}\frac{1}{(l-q)^2-m_{\phi''}^2+i\epsilon} \frac{1}{\slashed p-\slashed l+ \slashed q - M_{B'} + i\epsilon}.
\end{equation}
Finally, coefficients $C_{\phi_1\phi_2}$ and $C_{\phi B}$ are tabulated in Table~\ref{Tab:CPP} and Table~\ref{Tab:CPB}, respectively.

\end{itemize}

\begin{table}[h!]
  \centering
  \setlength{\tabcolsep}{20pt}
  \begin{tabular}{cccccc}
     \hline\hline
     $C_{\phi B}^{(V)}$ & $p$ & $n$ & $\Lambda$ & $\Sigma^0$ & $\Sigma^+$ \\
     \hline
     $K^-$ & $2$ & 0 & 0 & 0 & 0 \\
     $\bar{K}^0$ & 0 & $1$ & 0 & 0 & 0 \\
     $\pi^0$ & 0 & 0 & $\frac{\sqrt{3}}{2}$ & $\frac{1}{2}$ & 0 \\
     $\eta$  & 0 & 0 & $\frac{3}{2}$ & $\frac{\sqrt{3}}{2}$ & 0 \\
     $\pi^-$ & 0 & 0 & 0 & 0 & $1$ \\
     \hline\hline
   \end{tabular}
  \caption{Coefficients $C_{\phi B}^{(V)}$ appearing in the CT contribution to the hadronic current [Eq.~(\ref{Eq:CT1})].}
  \label{Tab:PB1}
\end{table}

\begin{table}[h!]
  \centering
  \setlength{\tabcolsep}{6pt}
  \begin{tabular}{cccccc}
     \hline\hline
     $C_{\phi B}^{(A)}$ & $p$ & $n$ & $\Lambda$ & $\Sigma^0$ & $\Sigma^+$ \\
     \hline
     $K^-$ & $-2F$ & 0 & 0 & 0 & 0 \\
     $\bar{K}^0$ & 0 & $-(D+F)$ & 0 & 0 & 0 \\
     $\pi^0$ & 0 & 0 & $-\frac{1}{2\sqrt{3}}(D+3F)$ & $\frac{1}{2}(D-F)$ & 0 \\
     $\eta$  & 0 & 0 & $-\frac{1}{2}(D+3F)$ & $\frac{\sqrt{3}}{2}(D-F)$ & 0 \\
     $\pi^-$ & 0 & 0 & 0 & 0 & $D-F$ \\
     \hline\hline
   \end{tabular}
  \caption{Coefficients $C_{\phi B}^{(A)}$ appearing in the CT contribution to the hadronic current [Eq.~(\ref{Eq:CT2})].}
\label{Tab:PB2}
\end{table}

\begin{table}[h!]
  \centering
  \setlength{\tabcolsep}{20pt}
  \begin{tabular}{*{6}{c}}
    \hline\hline
    $C_{\phi_1\phi_2}$ & $K^-$ & $\bar{K}^0$ & $\pi^0$ & $\eta$ & $\pi^-$ \\
    \hline
    $\pi^0$ & $-\frac{1}{\sqrt{2}}$ & 0 & 0 & 0 & 0 \\
    $\eta$  & $-\sqrt{\frac{3}{2}}$ & 0 & 0 & 0 & 0 \\
    $\pi^+$ & 0 & $-1$ & 0 & 0 & 0 \\
    $K^+$ & 0 & 0 & $\frac{1}{\sqrt{2}}$ & $\sqrt{\frac{3}{2}}$ & 0 \\
    $K^0$ & 0 & 0 & 0 & 0 & $1$ \\
    \hline\hline
  \end{tabular}
  \caption{Coefficients $C_{\phi_1\phi_2}$ appearing in the MF contribution to the hadronic current [Eq.~(\ref{Eq:MF})].}
  \label{Tab:CPP}
\end{table}

\begin{table}[h!]
  \centering
  \setlength{\tabcolsep}{5pt}
  \begin{tabular}{cccccc}
    \hline\hline
    $C_{\phi B}$ & $p$ & $n$ & $\Lambda$ & $\Sigma^0$ & $\Sigma^+$ \\
    \hline
    $\pi^0$ & $D+F$ & 0 & 0 & 0 & 0 \\
    $\eta$  & -$\frac{1}{\sqrt{3}}(D-3F)$ & 0 & 0 & 0 & 0 \\
    $\pi^+$ & 0 & $\sqrt{2}(D+F)$ & 0 & 0 & 0 \\
    $K^+$ & 0 & 0 & $-\frac{1}{\sqrt{3}}(D+3F)$ & $D-F$ & 0 \\
    $K^0$ & 0 & 0 & 0 & 0 & $\sqrt{2}(D-F)$ \\
    \hline\hline
  \end{tabular}
  \caption{Coefficients $C_{\phi B}$ appearing in the MF contribution to the hadronic current [Eq.~(\ref{Eq:MF})].}
\label{Tab:CPB}
\end{table}

The hadronic current presented above does not take into account the $q^2$ dependence of the weak interaction vertices, which is poorly known. Following Ref.~\cite{manolok}, we have parametrized this dependence with a global dipole form factor  
\begin{equation}
\label{eq:FF}
F(q^2) = \left(1 - \frac{q^2}{M_F^2} \right)^{-2} 
\end{equation}
that multiplies all the terms in $H_\mu$. Up to SU(3) breaking effects, the value of the axial mass $M_F$ should be similar to the one in electromagnetic and axial nucleon form factors. Therefore, as in Refs.~\cite{RafiAlam:2010kf,manolok} we have adopted $M_F \simeq 1$~GeV, accepting an uncertainty of around 10~\%. 

\subsection{Non-relativistic reduction of the invariant amplitude}

Because we only focus on the small momenta of the $\phi B$ components creating the $\Lambda(1405)$, we can perform a  non relativistic reduction, which was also used in the description of the $\phi B$ amplitude in coupled channels of Ref.~\cite{angels}. For the CT we get

\begin{eqnarray}
  -it^{\mathrm{CT}(V)} &=& -\frac{1}{4F_\phi}(2G_FV_{us})L^{0} \left[ C_{\phi B}^{(V)} + \sum\limits_{\phi' B'} C_{\phi' B'}^{(V)} G'_{\phi'B'} T_{\phi'B'\rightarrow \phi B}\right],\nonumber\\
  -it^{\mathrm{CT}(A)} &=& +\frac{1}{4F_{\phi}}(2G_FV_{us}) (\vec{L}\cdot \vec\sigma) \left[C_{\phi B}^{(A)} + \sum\limits_{\phi' B'} C_{\phi' B'}^{(A)} G'_{\phi'B'} T_{\phi'B'\rightarrow \phi B}\right],
\end{eqnarray}
where the loop function, after removing the baryon negative energy part, becomes
\begin{eqnarray}
  G'_{\phi' B'} &=& \int\frac{d^3 l}{(2\pi)^3} \frac{1}{2\omega_{\phi'}(\vec{l})}\frac{M_{B'}}{E_{B'}(\vec{l})} \frac{1}{M_{\mathrm{inv}}-\omega_{\phi'}(\vec{l})-E_{B'}(\vec{l}) + i\epsilon} \,.
\end{eqnarray}

After the non relativistic reduction, the MF contributions can be written as
\begin{eqnarray}
  -it^\mathrm{MF} &=& \frac{1}{4\sqrt{2}F_{\phi}}(2G_FV_{us})\left\{\sum\limits_{\phi''}C_{\phi''\phi}~C_{\phi''B}~\vec{\sigma}\cdot\vec{q}~ \frac{L^0(2k'-q)^0 +\vec{L}\cdot\vec{q}}{(k'-q)^2-m_{\phi''}^2+i\epsilon} \right.\nonumber\\
  && \left. + \sum\limits_{\phi'\phi''B'}C_{\phi''\phi'}C_{\phi''B'} \left[\vec{L}\cdot\vec{\sigma} G^{(1)}_{\phi'\phi''B'} + (\vec{L}\cdot\vec{q}) (\vec{\sigma}\cdot\vec{q}) G_{\phi'\phi''B'}^{(2)}\right]\right\} \,,
\end{eqnarray}
where the loop functions are
\begin{eqnarray}
 G_{\phi\phi' B'}^{(1)} &=& \int\frac{d^3l}{(2\pi)^3} \frac{1}{\omega_{\phi'}(\vec{l})\omega_\phi(\vec{l}-\tilde{\vec{q}})}\frac{M_{B'}}{E_{B'}(\vec{l})} \frac{\vec{l}^2}{3} \left\{\left[\omega_\phi(\vec{l}-\tilde{\vec{q}}) + \omega_{\phi'}(\vec{l})\right]^2 \right.\nonumber\\
  && \left.+ \left[\omega_\phi(\vec{l}-\tilde{\vec{q}}) + \omega_{\phi'}(\vec{l})\right]\left[E_{B'}(\vec{l})-\tilde{p}^0\right] -\tilde{q}^0\omega_{\phi'}(\vec{l}) \right\}\nonumber\\
  &&\times \frac{1}{M_\mathrm{inv}-E_{B'}(\vec{l}) - \omega_{\phi'}(\vec{l}) + i\epsilon} \frac{1}{\tilde{p}^0-E_{B'}(\vec{l})-\omega_\phi(\vec{l}-\tilde{\vec{q}}) + i\epsilon } \nonumber\\ && \times\frac{1}{\tilde{q}^0+\omega_\phi(\vec{l}-\tilde{\vec{q}})+\omega_{\phi'}(\vec{l})-i\epsilon}
   \frac{1}{\omega_{\phi'}(\vec{l})-\tilde{q}^0 + \omega_\phi(\vec{l}-\tilde{\vec{q}}) - i\epsilon}\,,
\end{eqnarray}
and
\begin{eqnarray}
  G_{\phi\phi' B'}^{(2)} &=& \int\frac{d^3l}{(2\pi)^3} \frac{1}{2\omega_{\phi'}(\vec{l})\omega_{\phi}(\vec{l}-\tilde{\vec{q}})}\frac{M_{B'}}{E_{B'}(\vec{l})} \left\{\left[\omega_{\phi}(\vec{l}-\tilde{\vec{q}}) + \omega_{\phi'}(\vec{l})\right]^2 \right.\nonumber\\
  &&  \left. + \left[\omega_{\phi}(\vec{l}-\tilde{\vec{q}}) + \omega_{\phi'}(\vec{l})\right]\left[E_{B'}(\vec{l})-\tilde{p}^0\right] -\tilde{q}^0\omega_{\phi'}(\vec{l}) \right\}\nonumber\\
  &&\times \frac{1}{M_\mathrm{inv}-E_{B'}(\vec{l}) - \omega_{\phi'}(\vec{l}) + i\epsilon} \frac{1}{\tilde{p}^0-E_{B'}(\vec{l})-\omega_{\phi}(\vec{l}-\tilde{\vec{q}}) + i\epsilon } \nonumber\\ &&\times\frac{1}{\tilde{q}^0+\omega_{\phi}(\vec{l}-\tilde{\vec{q}})+\omega_{\phi'}(\vec{l})-i\epsilon}
  \frac{1}{\omega_{\phi'}(\vec{l})-\tilde{q}^0 + \omega_\phi(\vec{l}-\tilde{\vec{q}}) - i\epsilon} \,.
\end{eqnarray}
The quantities with tilde are defined in the $\phi B$ CM frame.

\section{Results}~\label{SecIII}

\begin{figure}[h!]
  \centering
  \includegraphics[width=0.47\textwidth]{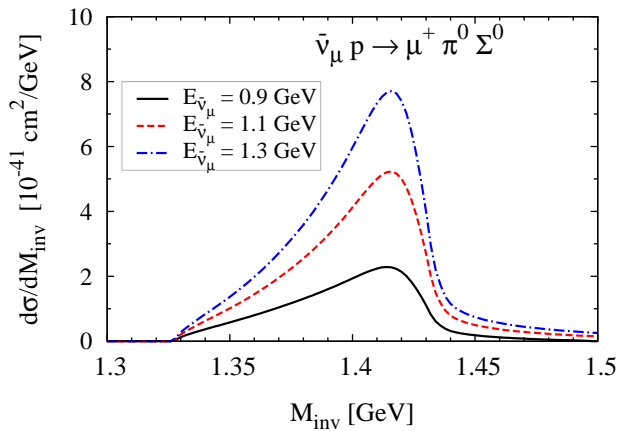}\\
  \caption{(color online). Differential cross section for the reaction $\bar{\nu}_\mu p \rightarrow \mu^+ \pi^0 \Sigma^0$ as a function of the invariant mass $M_\mathrm{inv}$ of the final meson baryon system for three different incident antineutrino energies.}
  \label{Fig:PSDSDM}
\end{figure}

\begin{figure}[h!]
  \centering
  \includegraphics[width=0.47\textwidth]{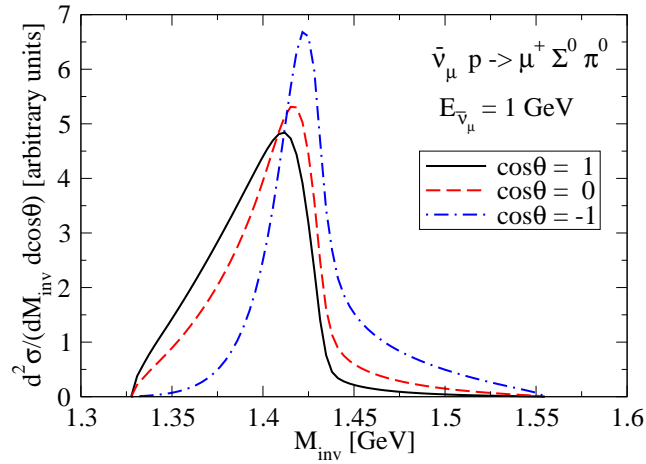}\\
  \caption{(color online). Area normalized double differential cross section for $\bar{\nu}_\mu p \rightarrow \mu^+ \pi^0 \Sigma^0$ at $E_{\bar{\nu}_{\mu}}=1$~GeV, as a function of $M_\mathrm{inv}$ for three different values of the angle ($\theta$) between the incoming neutrino and the outgoing muon in the reaction CM frame.}
  \label{Fig:d2s}
\end{figure}

Throughout this section, the results are presented for the muon flavor $l = \mu$. 
The $\Lambda(1405)$ can be observed in the invariant mass distribution of $\pi\Sigma$ pairs 
that has its threshold below the peak of the $\Lambda(1405)$ states. The cleanest signal for $I=0$ $\Lambda(1405)$ production appears in the $\pi^0\Sigma^0$ channel because $I=1$ is not allowed.
In Fig.~\ref{Fig:PSDSDM}, we show $d\sigma/dM_\mathrm{inv}$ for $\pi^0\Sigma^0$ production at three different laboratory energies, $E_{\bar\nu}=900$, $1100$, and $1300$~MeV. We can clearly see the resonant shape of
the $\Lambda(1405)$ at all the energies. Note that, in spite of the two poles, there is a single 
peak. This is common to all the reactions, with the exception of electroproduction~\cite{electro}, 
where the data are still relatively poor. Only the different weight of the two poles makes the peak
appear at different energies in different processes. In the present case the distribution peaks
around $1420$ MeV indicating that there is more weight from the pole at $1420$ MeV 
or, in other words, that the $\Lambda(1405)$ production induced by the $K^-p$ is dominant.
To gain further insight into the interplay of the two poles of the $\Lambda(1405)$ resonance in this reaction, we have looked at the line shapes of the double differential cross section $d^2\sigma/\left(dM_\mathrm{inv} d\cos{\theta}\right)$ for different values of the $\theta$ angle between the initial $\bar{\nu}_\mu$ and the final $\mu^+$ in the $\bar{\nu} p$ CM frame (Fig.~\ref{Fig:d2s}). When $\theta$ increases, so does $|q^2|$, and the form factor causes a reduction in the cross section. To compare the shapes we have normalized all curves to the same area by multiplying the cross section at $\cos{\theta} = 0 (-1)$ by 3.4(14). In the backward direction, the distribution clearly resembles a single Breit-Wigner with a mass and a width remarkably close to the values of the heavier pole of the $\Lambda(1405)$. It is this pole that appears dominant at this kinematics. As $\theta$ decreases, the presence of the lighter state becomes more evident with larger strength accumulating below the peak, which is shifted towards smaller invariant masses. The line shape becomes asymmetric but the second state never shows up as a peak in the cross section.    

\begin{figure}[h!]
  \centering
  \includegraphics[width=0.47\textwidth]{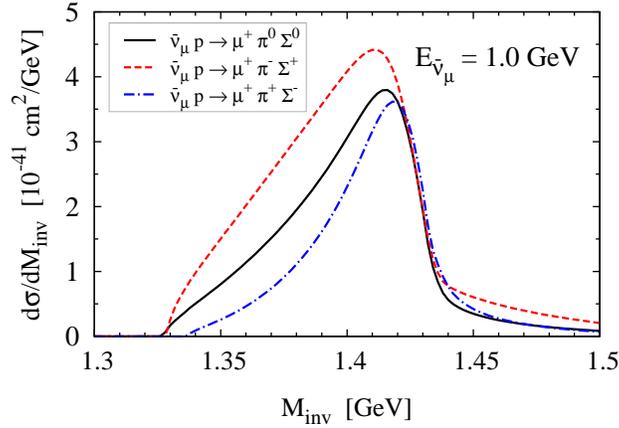}\\
  \caption{(color online). Invariant mass distribution for the three charge channels: $\pi^0\Sigma^0$ (solid line), $\pi^-\Sigma^+$ (dashed line) and $\pi^+\Sigma^-$ (dot-dashed line). The incident antineutrino energy is $E_{\bar{\nu}_{\mu}}=1$~GeV.}
  \label{Fig:PSDSDM3}
\end{figure}

It is also very interesting to consider $d\sigma/dM_\mathrm{inv}$ for the three charged channels 
$\pi^0\Sigma^0$, $\pi^+\Sigma^-$ and $\pi^-\Sigma^+$. This is shown in Fig.~\ref{Fig:PSDSDM3}. 
The peak position for the different reactions is slightly shifted, but the largest differences are 
present below the maxima. This is due to the contribution of an $I=1$ amplitude which adds constructively or destructively depending on the channel~\cite{rocados}. 
It was also shown in Ref.~\cite{rocados} that $\Lambda(1405)$ photoproduction data hint to a possible $I=1$ state around $1400$ MeV, which appears in some approaches~\cite{ollerulf} but is at a border line in others~\cite{cola}. In the work of Refs.~\cite{Wu:2009nw,Gao:2010hy}, the existence of
such $I=1$ state is claimed from the study of the $K^-p\rightarrow \Lambda\pi^-\pi^+$ reaction. The large differences seen in the cross sections for the three $\pi\Sigma$ channels in the present reaction indicate that they are indeed rather sensitive to the $I=1$ amplitude and, thus, there is a potential for the extraction of information on the possible $I=1$ state.

\begin{figure}[h!]
  \centering
  \includegraphics[width=0.47\textwidth]{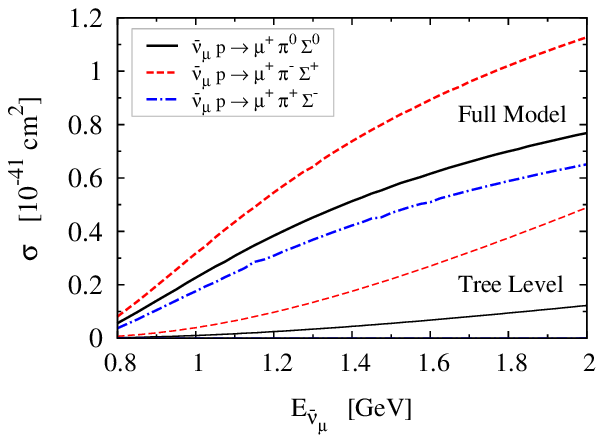}
  \caption{(color online). Cross sections as a function of the antineutrino energy for the three $\bar{\nu}_\mu p \rightarrow \mu^+ \pi \Sigma$ reaction channels. The three upper curves have been obtained with the full model while the two lower ones with tree level contributions alone. The later is absent for the $\pi^+ \Sigma^-$ channel.}  
  \label{Fig:PSFT}
\end{figure}

In Fig.~\ref{Fig:PSFT}, we show now the integrated cross sections for $\pi^0\Sigma^0$, $\pi^-\Sigma^+$, and $\pi^+\Sigma^-$ production. We observe a steady growth of the cross sections with the antineutrino energy.
These  cross sections are largely driven by the $\Lambda(1405)$ resonance. Indeed, in Fig.~\ref{Fig:PSFT}, both tree level and full model cross sections are shown. We observe that the contribution of the meson-baryon rescattering has a drastic effect in the results. The case of the $\pi^+\Sigma^-$ channel is the most spectacular because the tree level contribution is exactly zero. 

\begin{figure}[h!]
  \includegraphics[width=0.47\textwidth]{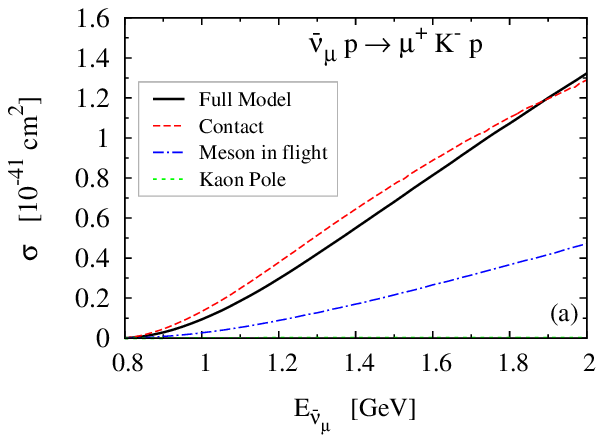}
  \includegraphics[width=0.47\textwidth]{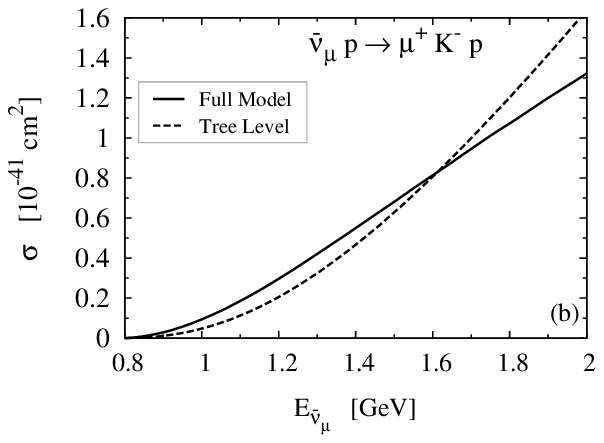}
  \caption{(color online). Integrated cross section for the $\bar{\nu}_\mu p \rightarrow \mu^+ K^- p$ reaction as a function of the antineutrino energy. Left panel: contribution of the different terms to the full model result. The KP contribution is negligible and cannot be discerned in the plot. Right panel: comparison between the full model and tree level calculations.}
  \label{Fig:KPS}
\end{figure}

\begin{figure}[h!]
  \includegraphics[width=0.47\textwidth]{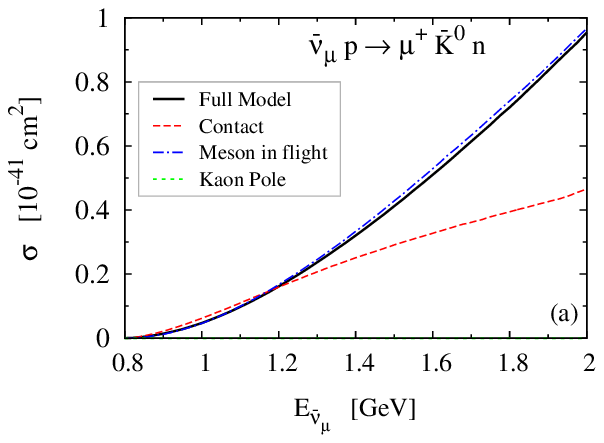}
  \includegraphics[width=0.47\textwidth]{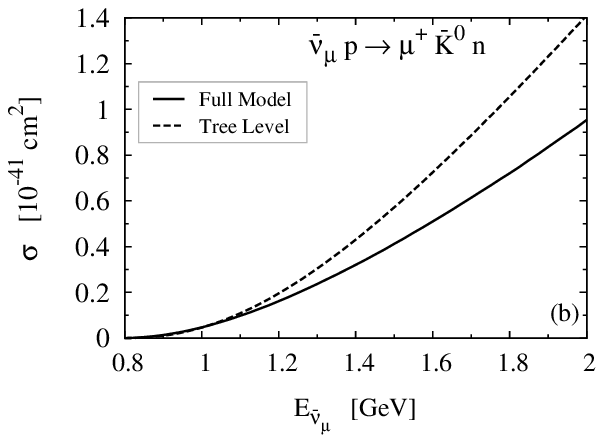}
  \caption{(color online). Integrated cross section for the $\bar{\nu}_\mu p \rightarrow \mu^+ \bar{K}^0 n$ reaction. The line styles have the same meanings as in Fig.~\ref{Fig:KPS}.}
  \label{Fig:KNS}
\end{figure}

We have also investigated the $\bar K$-nucleon production reactions. Note that in this case the threshold energies, $\sqrt{s}=m_{K^-} + M_p=1430$~MeV  and $m_{\bar K^0} + M_n =1437$~MeV, are already above the $\Lambda(1405)$ peak. Thus, we do not plot $d\sigma/dM_\mathrm{inv}$ in this case and show only the integrated cross section as a function of energy. These are shown in Fig.~\ref{Fig:KPS} for $K^-p$ and in Fig.~\ref{Fig:KNS} for $\bar{K}^0n$. 
As can be seen in the right panels of Figs.~\ref{Fig:KPS},\ref{Fig:KNS}, unlike the $\pi \Sigma$ production case, the cross section is not increased by the resonance.  On the contrary, the fast fall down of $d\sigma/dM_\mathrm{inv}$ close to the $K^-p$ threshold, seen in Fig.~\ref{Fig:PSDSDM} for $\pi\Sigma$, reflects the similar trend of the $t$ matrix which is common to all the channels. This affects the $\bar K$-nucleon production cross sections, most noticeably for $\bar{K}^0n$, the channel with a larger threshold. These unitarization effects were absent in the calculations reported in Ref.~\cite{manolok}. There are other differences between the present study and the one of Ref.~\cite{manolok}. First, here we have used the average $F_\phi = 1.15 f_\pi$, for consistency with the value taken in the study of $\phi B$ scattering~\cite{angels} (see Sec.~\ref{subsec:UChPT}), instead of $F_\phi = f_\pi$ in Ref.~\cite{manolok}. This leads to little smaller cross section with respect to those of Ref.~\cite{manolok}. Furthermore, the $p$-wave contributions considered in Ref.~\cite{manolok} but not here make the cross sections bigger as one departs from threshold. Finally, the non relativistic approximation becomes poorer for the higher energy and momentum transfers that can be probed as the reaction energy increases.  As an example, the CT contribution here is about 30\% lower than in Ref.~\cite{manolok} at $E_{\bar{\nu}}=1200$~MeV and about 40-45\% smaller at $E_{\bar{\nu}}=2000$~MeV (after correcting for $F_\phi$). For better precision, one should restrict to smaller antineutrino energies or implement kinematic cuts to keep $q^0$ and $|\vec{q}|$ small compared to the nucleon mass.

In the $K^-p$ channel, the largest contribution arises from the CT mechanism (left panel of Fig.~\ref{Fig:KPS}), in line with Fig.~3 of Ref.~\cite{manolok}. In the $\bar{K}^0 n$ channel, instead, the MF contribution becomes increasingly larger than the CT above $E_{\bar{\nu}}=1200$~MeV (left panel of Fig.~\ref{Fig:KNS}), in variance with Fig.~5 of Ref.~\cite{manolok}. Nevertheless, it should be mentioned that our predictions for KP, CT and MF terms converge to those of Ref.~\cite{manolok} in the heavy-nucleon limit.

\subsection{$\mathbf{\Lambda(1405)}$ production at MINER$\mathbf{\nu}$A}

One of the goals of the MINER$\nu$A experiment is to study weak strangeness production~\cite{Drakoulakos:2004gn}. It is therefore important to obtain the number of events in which the $\Lambda(1405)$ resonance is primarily produced during the antineutrino run. Let us consider the process $\bar{\nu}_\mu p \rightarrow \mu^+ \pi \Sigma$. The number of events for a given invariant mas of the $\pi \Sigma$ pair is 
\begin{equation}
\frac{d N}{d M_{\mathrm{inv}}} = N_{\mathrm{POT}} f M N_A \int dE_{\bar\nu} \phi(E_{\bar\nu}) \frac{d \sigma_{\pi \Sigma}}{d M_{\mathrm{inv}}}(E_{\bar\nu}) \,.
\end{equation}
The differential cross section is averaged over the antineutrino flux $\phi(E_{\bar\nu})$. The flux prediction, in units of $\bar\nu/cm^2/\mathrm{POT}$, for the low-energy configuration is taken from Table V of Ref.~\cite{Higuera:2014azj}. The present estimate corresponds to a number of protons on target of $N_{\mathrm{POT}} =  2.01 \times 10^{20}$ in $\bar\nu$ mode, neglecting the small $\bar\nu_e$ component in the beam of muon antineutrinos. Although the MINER$\nu$A detector is made of different materials, here we consider only the scintillator (CH). In this case the proton fraction $f=(1+6)/(1+12)$. One should recall that $\pi \Sigma$ pairs can also be produced on neutrons but, in this case, the pair has negative charge, not leading to  $\Lambda(1405)$ excitation. The scintillator mass is $M = 0.45 M_1 + 0.55 M_2$, with $M_1 =2.84 \times 10^6$ and  $M_2 =5.47 \times 10^6$ grams, to take into account that 45\% of the $\bar\nu$ data were taken during the construction time, using a reduced fiducial volume~\cite{Higuera:2014azj}. Finally, $N_A$ denotes the Avogadro number. 

\begin{figure}[h!]
  \includegraphics[width=0.47\textwidth]{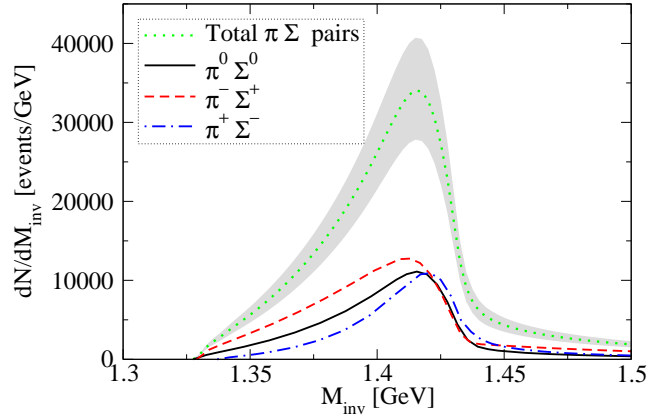}
  \caption{(color online). Invariant mass distribution of $\pi \Sigma$ events, primarily produced at the  MINER$\nu$A scintillator detector. The grey band corresponds to a 10\% error in the form factor parameter $M_F$.}  
  \label{Fig:Minerva}
\end{figure}

The event distributions for $\pi^0 \Sigma^0$, $\pi^- \Sigma^+$ and $\pi^+ \Sigma^-$ pairs and their sum, in the region of the  $\Lambda(1405)$ resonance, are shown in Fig.~\ref{Fig:Minerva}. At $q^2=0$, the largest invariant mass shown in Fig.~\ref{Fig:Minerva}, corresponds to a still moderate $\tilde{q}^0 = 456$~MeV, regardless of the antineutrino energy which can be high at  MINER$\nu$A ($\langle E_{\bar\nu} \rangle \sim 3.5$~GeV). For negative values of $q^2$, the largest $\tilde{q}^0$ can be larger, and even more so $|\tilde{\vec{q}}|$. On the other hand, the cross section for these $q^2$ is suppressed by poorly known vector and axial form factors, which have been accounted here with the global form factor of Eq.~\ref{eq:FF}. The uncertainty in the number of events at non-zero $q^2$, accounted by a 10\% error in $M_F$, is represented by the band in Fig.~\ref{Fig:Minerva}. 
By integrating the distributions in Fig.~\ref{Fig:Minerva}, one finds the following numbers of events: $N_{\pi^0 \Sigma^0} = 612^{+120}_{-112}$,  $N_{\pi^+ \Sigma^-} = 517^{+100}_{-94}$,  $N_{\pi^- \Sigma^+} = 838^{+163}_{-153}$. All in all, we predict about 2000 $\pi \Sigma$ pairs coming predominantly from $\Lambda(1405)$ decay.   

Modern neutrino experiments, including MINER$\nu$A, have detectors with nuclear targets. Nuclear effects, not considered in the present study, play an important role. It has been shown that strangeness can be abundantly produced in secondary collisions~\cite{Lalakulich:2012gm}. The events predicted above correspond to $\Lambda(1405)$ excitation in primary $\bar\nu N$ collisions but the actual signal will be different. The invariant mass of the outgoing $\pi \Sigma$ gets distorted by final state interactions with other nucleons in the nucleus~\footnote{These genuinely nuclear processes should not be confused with the unitarization mechanisms at the nucleon level that generate the $\Lambda(1405)$ dynamically, as discussed above.}; the composition of the final state can change because of pion absorption and other inelastic processes like $\pi \, N \rightarrow K \, Y$, $\Sigma \, N \rightarrow N \, N \, \bar K$ and others. In the same way, the $\Lambda(1405)$ can be produced in secondary $\bar K N$ scattering. This dynamics requires a more detailed investigation to find specific indications of $\Lambda(1405)$ production in $\bar\nu$-nucleus collisions. Yet, as it happens in photonuclear reactions in nuclei, even if secondary collisions distort the resonance signal, there is still a sizeable fraction of events not affected by them. These events mostly come from primary interactions taking place in the back of the nucleus with respect to the direction of the three-momentum transfer $\vec{q}$ in the Laboratory frame. Therefore, a signal from the primary collisions can be observed in these reactions. This is the case in $\Delta(1232)$~\cite{carrasco,carrasco1}  and $\omega$~\cite{metag} photoproduction.

\section{Conclusions}~\label{SecIV}

We have studied $\Lambda(1405)$ production induced by antineutrinos, the first calculation of this sort. For this purpose we have combined elements of chiral perturbation theory in the presence of weak external fields with unitarization techniques in coupled channels. The $\Lambda(1405)$, consisting actually of two states, is generated through the multiple scattering of meson-baryon coupled channels with a kernel provided by the chiral Lagrangians. It can only be observed in the $\pi\Sigma$ final state, most cleanly in the $\pi^0\Sigma^0$ channel which has only $I=0$.  As in most reactions, the $\Lambda(1405)$ appears as a single highly asymmetric peak in the  $\pi \Sigma$ invariant mass distribution. The line shapes at different angles between the incoming $\bar\nu$ and the outgoing lepton in the reaction CM frame indicate that the process at backward angles is dominated by a state with mass and width of around 1420 and 40~MeV, respectively. As the angle decreases, the lighter states becomes increasingly more important.  

The $\pi^+\Sigma^-$ and $\pi^-\Sigma^+$ channels also contain an $I=1$ amplitude, where a possible resonance might be present according to some studies. This amplitude is responsible for large differences in the shapes of the $\pi \Sigma$ invariant mass distributions below the maximum for the three charge channels. Therefore, a combined study of $\pi^0\Sigma^0$, $\pi^+\Sigma^-$ and $\pi^-\Sigma^+$ production induced by antineutrinos could provide useful information about this hypothetical $I=1$ state.

We have also evaluated the integrated cross sections for $\bar{\nu}_\mu p \rightarrow \mu^+ \pi \Sigma$ as a function of the antineutrino energy. These are much larger than the corresponding tree level results due to the $\Lambda(1405)$ excitation. We should note that the tree level is relatively more important for the $\bar{K} N$ final state because the latter is above the $\Lambda(1405)$. In this case, unitarization does not cause an enhancement of the cross section. One rather observes a reduction in the $\bar{K}^0n$ channel, which has the largest threshold.   

We have obtained that the number of events in which the $\Lambda(1405)$ is excited in primary $\bar\nu_\mu p$ collisions at the scintillator detector of the MINER$\nu$A experiment, in the antineutrino run, is of the order of 2000. It is large enough to conclude that $\Lambda(1405)$ production has a sizable impact in the scattering dynamics leading to antineutrino detection, and should be taken into account in future evolutions of neutrino event generators.  

Several open questions in the physics of (anti)neutrino interactions with matter call for new measurements of (anti)neutrino cross sections on proton and hydrogen targets~\cite{Alvarez-Ruso:2014bla}. Such experiments with antineutrinos would also provide a more complete understanding of the $\Lambda(1405)$ properties.

\begin{acknowledgments}
LAR wishes to thank J. Morfin for a useful discussion about the flux and event rate at the MINER$\nu$A experiment. The work of LAR has been partially supported by the U.S. Department of Energy, Office of Science, Office of High Energy Physics, through the Fermilab Intensity Frontier Fellows Program. He gratefully acknowledges the hospitality during his stay at Fermilab.  X.-L.R thanks Prof. Jie Meng and Prof. Li-Sheng Geng for useful discussions. He acknowledges support from the Innovation Foundation of Beihang University for Ph.D. Graduates, the National Natural Science Foundation of China under Grant No. 11375024 and a fellowship from the China Scholarship Council. This work has been partly supported by the Spanish Ministerio de Econom\'ia y Competitividad and European FEDER funds under the contract number FIS2011-28853-C02-01 and FIS2011-28853-C02-02, and the Generalitat Valenciana in the program Prometeo II, 2014/068.
We acknowledge the support of European Community-Research Infrastructure Integrating Activity Study of Strongly Interacting Matter (acronym HadronPhysics3, Grant Agreement n. 283286) under the Seventh Framework Program of EU.
\end{acknowledgments}

\end{document}